\documentclass{PoS}

\usepackage{graphicx}
\usepackage{amsmath,amssymb,amsfonts}
\usepackage{subfigure}
\usepackage{listings}
\usepackage{cite}

\newcommand{\GOSAM}{{\textsc{Go\-Sam}}}

\newcommand{\QGRAF}{{\texttt{QGRAF}}}
\newcommand{\FORM}{{\texttt{FORM}}}
\newcommand{\SPINNEY}{{\texttt{spin\-ney}}}
\newcommand{\HAGGIES}{{\texttt{hag\-gies}}}
\newcommand{\SAMURAI}{{\textsc{Sa\-mu\-rai}}}

\newcommand{\bea}{\begin{eqnarray}}
\newcommand{\eea}{\end{eqnarray}\noindent}

\newcommand{\bcen}{\begin{center}}
\newcommand{\ecen}{\end{center}}
\def\url#1{\texttt{#1}}

\def\Gosam{{{\sc GoSam}}}
\def\gosam{{{\sc GoSam}}}

\def\samurai{{{\sc samurai}}}
\def\Sherpa{{{\sc Sherpa}}}

\def\C++{{{\sc c++}}}

\def\QCDLoop{{{\sc QCDLoop}}}
\def\OneLoop{{{\sc OneLoop}}}
\def\Golem{{{\sc Golem95C}}}

\def\Ninja{{{\sc Ninja}}}

\def\QCDLoop{{{\sc QCDLoop}}}

\title{NLO QCD Production of Higgs plus jets with GoSam}

\ShortTitle{NLO QCD Production of Higgs boson plus jets with GoSam}

\author{G.~Cullen\\
Deutsches Elektronen-Synchrotron DESY, Platanenallee 6, 15738 Zeuthen, Germany\\
E-mail: \email{gavin.cullen@desy.de}}

\author{H.~van Deurzen, N.~Greiner, G.~Heinrich,  G.~Luisoni, E.~Mirabella, 
T.~Peraro, J.~Reichel, J.~Schlenk, J.F. von Soden-Fraunhofen\\
Max Planck Institute for Physics, F\"ohringer Ring 6, 80805 Munich, Germany\\
E-mail: \email{\{hdeurzen, greiner, gudrun, luisonig, mirabell, peraro, joscha, jschlenk, jfsoden\}@mpp.mpg.de}}

\author{P.~Mastrolia\\
Max Planck Institute for Physics, F\"ohringer Ring 6, 80805 Munich, Germany; \\ Dipartimento di Fisica e Astronomia, Universit\`a di Padova, and INFN Sezione di Padova, 
via Marzolo 8, 35131 Padova, Italy\\
 E-mail: \email{ppaolo@mpp.mpg.de}}

\author{\speaker{G.~Ossola}\\
Physics Department, New York City College of Technology, The
City University of New York,
300 Jay Street Brooklyn, NY 11201, USA; \\
The Graduate School and University Center, The City University of New York,
365 Fifth Avenue, New York, NY 10016, USA\\
 E-mail: \email{gossola@citytech.cuny.edu}}

\author{F.~Tramontano\\
Dipartimento di Scienze Fisiche, Universit\`a degli studi di Napoli and INFN, Sezione di Napoli, 
80125 Napoli, Italy\\
E-mail: \email{francesco.tramontano@cern.ch}}

\abstract{
After reviewing the main features of the {\gosam} framework for automated one-loop calculations, we present a selection of recent phenomenological results obtained with it. In particular, we focus on the recent calculation of NLO QCD corrections to the production of a Higgs boson in conjunction with jets at the LHC.
}

\FullConference{The European Physical Society Conference on High Energy Physics \\
                 18-24 July, 2013\\
                 Stockholm, Sweden}

\begin{document}

\section{Introduction}

The large amount of data accumulated by the experimental 
collaborations at the Large Hadron Collider (LHC) allowed for a very 
detailed investigation of the Standard Model (SM) of particle physics.
Moreover, the discovery of a Higgs boson with mass of about 126 GeV~\cite{Aad:2012tfa
} finally confirmed the validity of the electroweak symmetry breaking mechanism~\cite{Englert:1964et
}. 

In all these analyses, for example to further study the properties of the recently discovered Higgs boson, theory predictions play a fundamental role. They are not only needed for the signal, but also for the modeling of the relevant background processes, which share similar experimental signatures. Further, precise theory predictions are important in order to constrain model parameters in the event that a signal of New Physics is detected. 
Since leading-order (LO) results are affected by large uncertainties, theory predictions are not reliable without accounting for higher orders. Therefore, it is of primary interest to provide theoretical tools which are able to perform the comparison of LHC data to theory  at NLO accuracy.

In the past few years, the progress in the automation of NLO calculations for multi-particle final states has been tremendous and led to the so-called ``NLO revolution''~\cite{Salam:2011bj
}. 
Several automated frameworks for one-loop calculations~\cite{Berger:2008sj
} have been presented, which are based on various new theoretical developments~\cite{Ellis:2011cr
}. It is fascinating to witness the number and quality of advanced automated NLO calculations that have been performed with different techniques. 

In this presentation, we review the main features of the {\gosam} framework~\cite{Cullen:2011ac} for the automated computation of one-loop amplitudes. {\gosam} has been recently employed in several calculations at NLO QCD accuracy~\cite{Greiner:2011mp,Greiner:2012im, vanDeurzen:2013rv, Gehrmann:2013aga,Gehrmann:2013bga, Cullen:2013saa, vanDeurzen:2013xla, Dolan:2013rja} related to signal and backgrounds for Higgs boson production, as well as in the context of Beyond Standard Model (BSM) scenarios~\cite{Cullen:2012eh,Greiner:2013gca} and electroweak studies~\cite{Chiesa:2013yma
}, and has been successfully interfaced with Monte Carlo programs to merge multiple NLO matrix elements with parton showers~\cite{Luisoni:2013cuh, Hoeche:2013mua}.

We also briefly describe a selection of recent phenomenological results obtained with {\gosam}, with particular attention to the recent calculations  of NLO QCD corrections to the production of Higgs boson in conjunction with jets at the LHC.

\section{The \GOSAM{} framework}

{\gosam} combines automated diagram generation and algebraic manipulation~\cite{Nogueira:1991ex, Vermaseren:2000nd, Reiter:2009ts, Cullen:2010jv} with integrand-level reduction
techniques~\cite{Ossola:2006us
}.
Amplitudes are generated via Feynman diagrams, using \QGRAF~\cite{Nogueira:1991ex}, \FORM~\cite{Vermaseren:2000nd}, \SPINNEY~\cite{Cullen:2010jv} and \HAGGIES~\cite{Reiter:2009ts}. 
The individual program tasks are managed by python scripts, so that the only task required from the user is the preparation of an input file in order to launch the generation of the source code and its compiling, without having to worry about  the internal details.
The input file contains specific information about: i) the {\it process}, such as a list of initial and
      final state particles, the order in the coupling constants, and the model;
ii) the {\it scheme} employed, such as
    the regularization and renormalization schemes;
iii) the {\it system} , such as paths to libraries or compiler options;
iv) optional information to control the code generation.

After the generation of all contributing diagrams, the virtual corrections are evaluated using the $d$-dimensional integrand-level reduction method, 
as implemented in \SAMURAI~\cite{Mastrolia:2010nb} library, which allows for the combined
determination of both cut-constructible and rational terms at once.
Alternatively, the tensorial decomposition provided by {\Golem}~\cite{Binoth:2008uq,Heinrich:2010ax,Cullen:2011kv}  is also available. Such reduction, which is numerically stable but more time consuming, is employed as a rescue system. After the reduction, all relevant master integrals can be computed by means of {\Golem}~\cite{Cullen:2011kv}, {\QCDLoop}~\cite{vanOldenborgh:1990yc
}, or {\OneLoop}~\cite{vanHameren:2010cp}.

As a novel approach to the integrand reduction, the method proposed in~\cite{Mastrolia:2012bu} allows for the extraction of all the coefficients in the integrand decomposition by performing a Laurent expansion, whenever the analytic form of the numerator function is known. This method has been implemented, within the {\gosam} framework, in the \C++ library {\Ninja}, showing an improvement in the computational performance, both in terms of speed and precision, with respect to the standard algorithms. More details are provided in the talk of T.~Peraro at this conference~\cite{TizianoEPS}. The new library has been recently employed in the evaluation of NLO QCD corrections to $p p \to t {\bar t} H j $~\cite{vanDeurzen:2013xla}.

\GOSAM{} can be used to generate and evaluate one-loop corrections in both QCD and electro-weak theory. 
Model files for BSM theories can be generated from a Universal FeynRules Output (\texttt{UFO})~\cite{Christensen:2008py
} or 
\texttt{LanHEP}~\cite{Semenov:2010qt} file.

\paragraph{Code development} \label{code}
New features have been recently implemented within \GOSAM{}, with respect  to the current public version. 
In order to deal with the complexity level of calculations such as $pp\to Hjjj$~\cite{Cullen:2013saa}, the {\gosam} code has been enhanced. On the one side, the generation algorithm has been improved by a more efficient diagrammatic layout: Feynman diagrams are grouped according to their
topologies, namely global numerators are constructed by combining
diagrams that have a common set, or subset, of denominators,
irrespectively of the specific particle content.  On the other side,
additional improvements in the performances of {\sc GoSam} have been
achieved by exploiting the optimized manipulation of polynomial
expressions available in {\sc Form 4.0}~\cite{Kuipers:2012rf}.  
The possibility of employing numerical polarization vectors and the option to sum diagrams sharing the same propagators algebraically during the generation of the code led to an enormous gain in code generation time and reduction of code size.

Concerning the amplitude reduction, aside from the already mentioned new integrand-level reduction via Laurent expansion~\cite{Mastrolia:2012bu}, {\sc GoSam} has been enhanced to reduce integrands that may exhibit numerators with rank larger than the number of the denominators. This is indeed the case in the presence of effective couplings~\cite{vanDeurzen:2013rv, Cullen:2013saa},  which appear in the large top-mass approximation, or when dealing with spin-2 particles~\cite{Greiner:2013gca}.
For these cases, within the context of integrand-reduction techniques, the parametrization of the residues at the
multiple-cut has to be extended and the decomposition of any one-loop amplitude acquires new master integrals~\cite{Mastrolia:2012bu}. The extended integrand decomposition
has been implemented in the \samurai\ library~\cite{Mastrolia:2012du}.

The new developments regarding the improved generation and
reduction algorithms will be publicly available in the \GOSAM\,2.0 release, which is currently in preparation.

\paragraph{Interfacing with MC and BLHA}

The computation of physical observables at NLO accuracy, such as cross sections and differential distributions, requires to combine the one-loop results for the virtual amplitudes obtained with \gosam{}, with other tools that can take care of the computation of the real emission contributions and of the subtraction terms, needed to control the cancellation of IR singularities.

This can be obtained by embedding the calculation of virtual corrections within a Monte Carlo framework (MC), 
 that can take care of the phase-space integration, and of the combination of the different pieces of the calculation. A table with a comprehensive list of \gosam{} interfaces with MC programs has been recently presented in~\cite{Cullen:2013cka}.

In order to facilitate the communication between the programs computing virtual one-loop amplitudes and the MC frameworks, a standard interface called the \emph{Binoth Les Houches Accord} (BLHA)~\cite{Binoth:2010xt
} has been designed. Within the BLHA, the interaction between the One-loop Program (OLP) and the Monte Carlo framework (MC) proceeds in two phases. During the first phase, called pre-runtime phase, the MC creates an order file, which contains information about the setup and the subprocesses needed
from the OLP in order to  perform the computation. The OLP reads the order file,  checks availability for each item, and  returns a contract file telling the MC what it can provide. In the second stage, the MC requires from the OLP  the values of the virtual one-loop amplitudes at specific phase-space points.  

\section{Higgs boson production in Gluon Fusion}
 
At the LHC, the dominant Higgs production mechanism proceeds
via gluon fusion (GF),  where the coupling of the Higgs to
the gluons is mediated by a heavy quark loop.
For this reason, the calculation of higher order corrections for the
GF production of a Higgs boson in association with jets has received a
lot of attention in the theory community over the past
decade~\cite{Dittmaier:2011ti
}.

\begin{figure}[h]
 \hspace{1.0cm} \subfigure[]{\includegraphics[width=6cm]{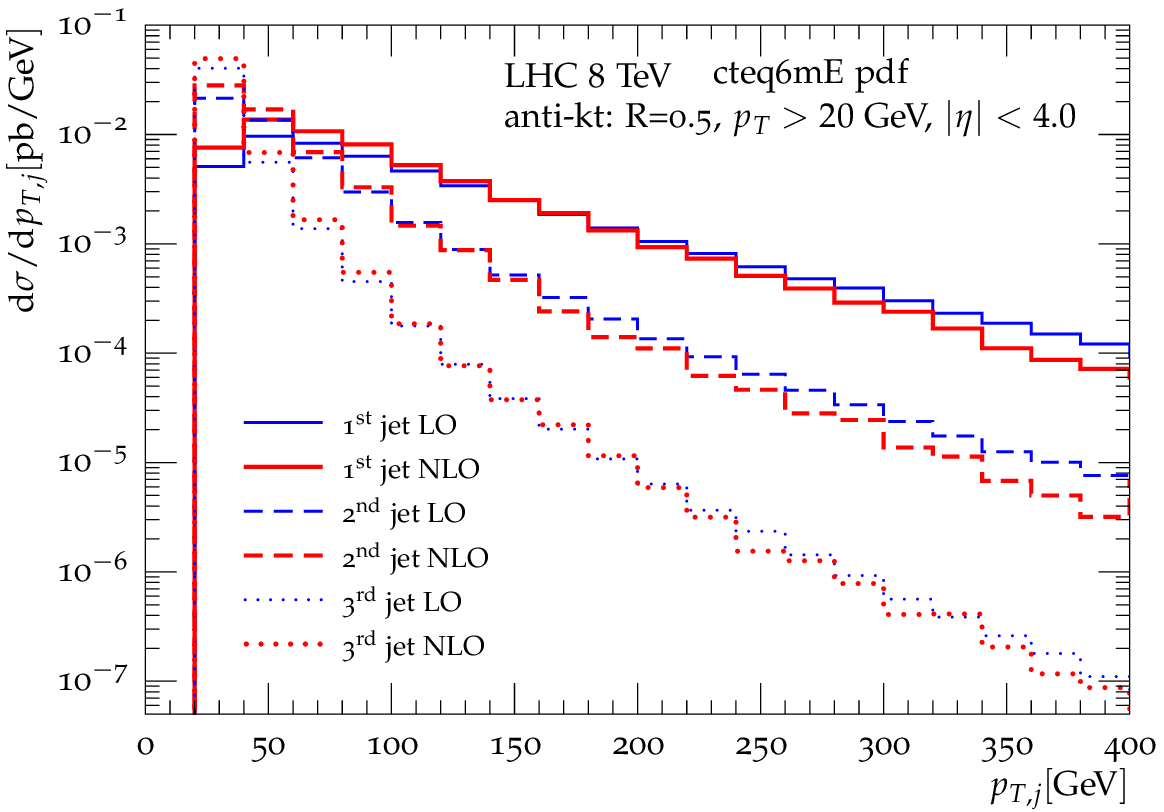}} \hspace{1.0cm}
\subfigure[]{\includegraphics[width=6cm]{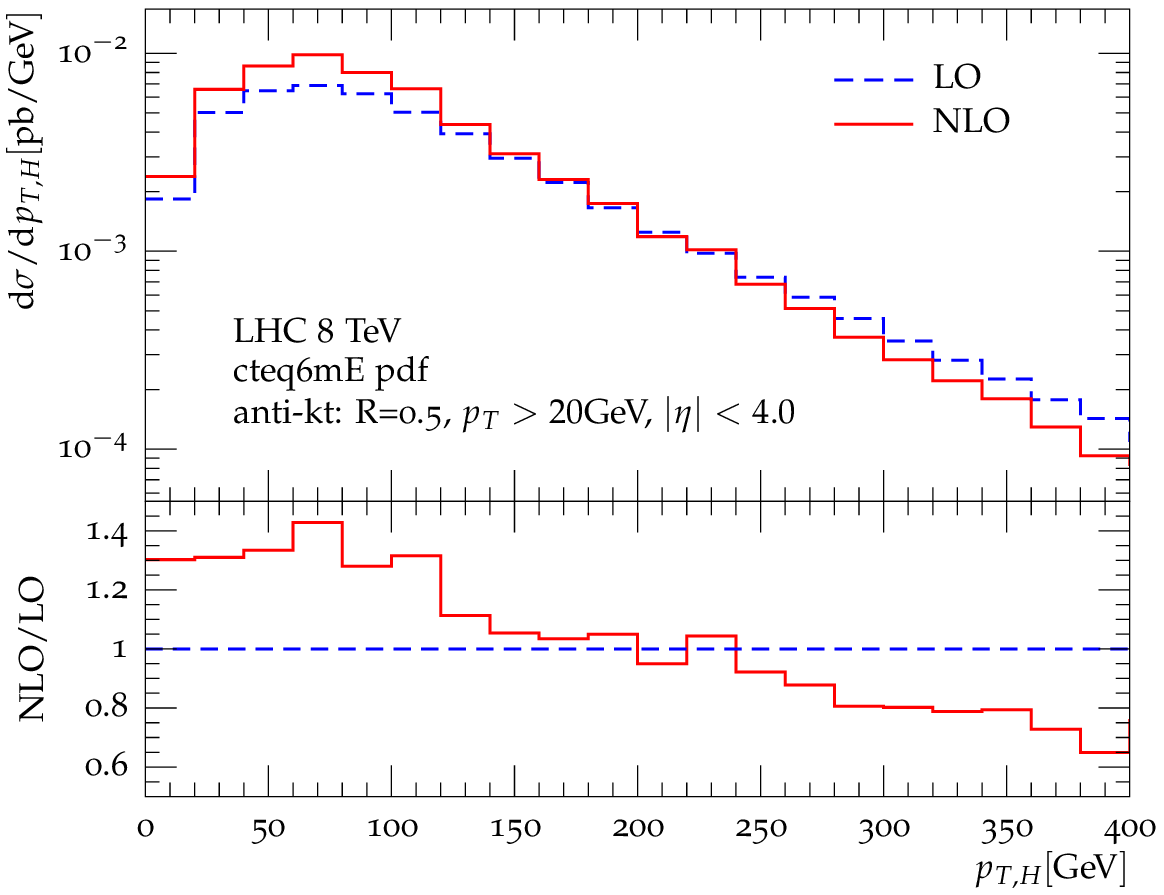}} 
\caption{ $pp \to H+3 j$ in GF for the LHC
at 8 TeV: (a) transverse momentum distributions of the leading jets, (b) transverse momentum distribution of the
  Higgs boson.}
\label{fig:H3j}
\end{figure}

The developments in \GOSAM{}, described in Section~\ref{code}, allowed to compute
the NLO QCD corrections to the production of $H+2$
jets~\cite{vanDeurzen:2013rv} and, for the first time, also $H+3$ jets~\cite{Cullen:2013saa} in
GF (in the large top-mass limit). 
While a fully automated BLHA interface between \GOSAM{}  and \Sherpa{}~\cite{Gleisberg:2008ta} has 
been used for $pp \to H+2 j$, the complexity of the integration for the process $pp \to H+3 j$ forced us to employ a hybrid setup which combines \GOSAM{}, \Sherpa{} and the MadDipole/Madgraph4/MadEvent framework~\cite{Frederix:2008hu
}. 
This calculation is indeed challenging both on the side of
real-emission contributions and of the virtual corrections, which alone
involve more than $10,000$ one-loop Feynman diagrams with up to
rank-seven hexagons.

In the calculation of $H+3$ jets the cteq6L1 and cteq6mE
parton-distribution functions were used for LO and NLO respectively,
and a minimal set of cuts based on the anti-$k_T$ jet algorithm with
$R=0.5$, $p_{T,min}>20$ GeV and $\left|\eta\right|<4.0$ was applied.
Figure~\ref{fig:H3j} shows the $p_T$
distributions of the three jets and of the Higgs boson,
respectively. The NLO corrections enhance all distributions for $p_T$
values lower than $150-200$ GeV, whereas their contribution is
negative at higher $p_T$. 

This study also shows that the virtual contributions for $pp \to Hjjj$
generated by {\Gosam} is ready to be paired with available Monte
Carlo programs to aim at further phenomenological studies.

\section{Other Phenomenological results}

\paragraph{Diphotons+jets}

\GOSAM{} in combination with MadDipole/MadGraph4/MadEvent
 has been used to calculate the NLO QCD corrections to 
$pp\to\gamma\gamma +1$\,jet\,\cite{Gehrmann:2013aga} and $pp\to\gamma\gamma +2$\,jets\,\cite{Gehrmann:2013bga}, where the former also includes the fragmentation component.
This calculation allowed for a first reliable prediction of the absolute normalization of this process, and demonstrated that the shape of important kinematical distributions is modified by higher-order effects.

\paragraph{Beyond the Standard Model}
\GOSAM{} has been used to calculate the NLO Susy-QCD corrections to the production of 
a pair of the lightest neutralinos plus one jet at the LHC at $8$\,TeV, which appears as a 
monojet signature in combination with missing energy. 
All non-resonant diagrams have been fully included, namely without using the assumption that production and decay factorize. 
We observe that the NLO corrections to the missing transverse energy are large, mainly due to additional channels opening up at NLO.
The detailed setup can be found in~\cite{Cullen:2012eh}.

Another recent BSM result obtained with \GOSAM{}+MadDipole/MadGraph4/MadEvent 
is the calculation of NLO QCD corrections to the production of a graviton in association with one jet~\cite{Greiner:2013gca},  where the  graviton decays into  a photon pair,   
within ADD models of large extra dimensions~\cite{ArkaniHamed:1998rs
}.
The calculation is quite complicated due to the tensor structure 
introduced by spin-2 particles, and the non-standard propagator of the graviton, 
coming from the summation over Kaluza-Klein modes.
It is interesting to notice that the $K$-factors of the invariant mass distribution of the photon pair emitted in the decay of the graviton are not uniform. Since the latter is used to derive exclusion limits, it is advisable to take into account the effect of NLO corrections. For details we refer to \cite{Greiner:2013gca}.

\paragraph{Higgs+Vector Boson+jet} {\Gosam} was interfaced with the POWHEG BOX to compute the associate production of a Higgs boson, a vector boson, and one jet~\cite{Luisoni:2013cuh}. In this calculation,  the improved MiNLO procedure~\cite{Hamilton:2012rf} was used to obtain NLO accurate predictions also when the jet is not resolved. Using this interface the generation of any process is fully automated, except for the construction of the Born phase space. 

\paragraph{NLO QCD corrections to $p p \to t {\bar t} Hj $} The production rate for a Higgs boson associated with a top-antitop pair ($t \bar t H$) is particularly interesting to study the properties of the newly discovered Higgs boson, since it is directly proportional to the SM Yukawa coupling of the Higgs boson to the top quark. 
We recently presented the complete NLO QCD corrections to the process $ pp \to t \bar t H + 1$ jet  ($t \bar t H j$) at the LHC~\cite{vanDeurzen:2013xla}. The goal of the calculation was twofold.  On the one hand, it is important for the phenomenological analyses at the LHC, in particular for  the high-$p_T$ region, where the presence of the additional jet can be relevant. On the other hand, from the technical point of view, for the presence of two mass scales (Higgs boson and top quark) and internal massive particles, together with a high number of diagrams,  $p p \to t {\bar t} H j $ constitutes a challenge for many reduction algorithms. 
This calculation represents the first application of the novel reduction algorithm, implemented in the library {\Ninja},  based on integrand-level reduction via Laurent expansion~\cite{Mastrolia:2012bu}. 

\section{Conclusions}

\GOSAM{} is a flexible and widely applicable tool for the automated calculation of the virtual part of multi-particle scattering amplitudes at NLO accuracy. After interfacing it with MC programs, that can perform integration over phase space and combine the contributions coming from real emission and subtraction terms as well, total cross-sections and differential distributions can be easily obtained for a variety of processes of interest at the LHC. 

Boosted by state-of-the-art techniques for the reduction of the scattering amplitudes,  \GOSAM{} provides a reliable answer for  multi-leg amplitudes in the presence of massive internal and external legs and propagators, such as the production of a Higgs boson in conjunction with a top-quark pair, as well as in configurations with relatively high multiplicity, such as Higgs boson plus jets production.
While the \GOSAM{} code will be further improved, it will be interesting to observe whether the attempt of extending integrand-level techniques to higher orders~\cite{Mastrolia:2011pr
}  will succeed and provide a comparable level of automation, at least for the calculation of the virtual parts. 

Other challenges for the near future involve interfacing \GOSAM{} with MC programs for an automated generation of the full cross section including parton showering, and ultimately the production of codes and results to be used within experimental analyses. We believe that the amount of recent calculations that were produced with the \GOSAM{} framework shows, both in terms of stability and precision, that it is an ideal multi-purpose tool for studying the physics of the LHC. 

\subsection*{Acknowledgments}
The work of G.C. was supported by DFG
Sonderforschungsbereich Transregio 9, Computergest\"utzte Theoretische Teilchenphysik and
the support of the Research Executive Agency (REA)
of the European Union under the Grant Agreement number
PITN-GA-2010-264564 (LHCPhenoNet).
H.v.D., G.L., P.M., and T.P. are supported by the Alexander von
Humboldt Foundation, in the framework of the Sofja Kovaleskaja Award Project
``Advanced Mathematical Methods for Particle Physics'', endowed by the German
Federal Ministry of Education and Research.
The work of G.O. was supported in part by the National Science Foundation
under Grant PHY-1068550 and PSC-CUNY Award No. 65188-00 43. 
This research work benefited of computing
resources from the Rechenzentrum Garching and the the CTP cluster of the New York City
College of Technology. 


\providecommand{\newblock}{}


\begin{thebibliography}{10}
\expandafter\ifx\csname url\endcsname\relax
  \def\url#1{{\tt #1}}\fi
\expandafter\ifx\csname urlprefix\endcsname\relax\def\urlprefix{URL }\fi
\providecommand{\eprint}[2][]{\url{#2}}

\bibitem{Aad:2012tfa}
Aad G {\em et~al.\/} (ATLAS Collaboration) 2012 {\em Phys.Lett.\/} {\bf B716}
  1--29;
Chatrchyan S {\em et~al.\/} (CMS Collaboration) 2012 {\em Phys.Lett.\/} {\bf
  B716} 30--61

\bibitem{Englert:1964et}
Englert F and Brout R 1964 {\em Phys.Rev.Lett.\/} {\bf 13} 321--323;
Higgs P~W 1964 {\em Phys.Lett.\/} {\bf 12} 132--133

\bibitem{Salam:2011bj}
Salam G~P 2010 {\em PoS\/} {\bf ICHEP2010} 556;
Perret-Gallix D 2013 {\em J.Phys.Conf.Ser.\/} {\bf 454} 012051
 

\bibitem{Berger:2008sj}
Berger C, Bern Z, Dixon L, Febres~Cordero F, Forde D {\em et~al.\/} 2008 {\em
  Phys.Rev.\/} {\bf D78} 036003;
Bevilacqua G, Czakon M, Garzelli M, van Hameren A, Kardos A {\em et~al.\/} 2013
  {\em Comput.Phys.Commun.\/} {\bf 184} 986--997;
Hirschi V, Frederix R, Frixione S, Garzelli M~V, Maltoni F {\em et~al.\/} 2011
  {\em JHEP\/} {\bf 1105} 044;
Cascioli F, Maierhofer P and Pozzorini S 2012 {\em Phys.Rev.Lett.\/} {\bf 108}
  111601;
Agrawal S, Hahn T and Mirabella E 2012 {\em PoS\/} {\bf LL2012} 046;
Badger S, Biedermann B, Uwer P and Yundin V 2013 {\em Comput.Phys.Commun.\/}
  {\bf 184} 1981--1998;
Actis S, Denner A, Hofer L, Scharf A and Uccirati S 2013 {\em JHEP\/} {\bf
  1304} 037 

\bibitem{Cullen:2011ac}
Cullen G, Greiner N, Heinrich G, Luisoni G, Mastrolia P {\em et~al.\/} 2012
  {\em Eur.Phys.J.\/} {\bf C72} 1889 

\bibitem{Ellis:2011cr}
Ellis R~K, Kunszt Z, Melnikov K and Zanderighi G 2012 {\em Phys.Rept.\/} {\bf
  518} 141--250;  Bern Z, Dixon L and Kosower D 2007 {\em Annals Phys.\/}  {\bf 322} 1587;
Ossola G 2013  (\textit{Preprint} \eprint{1310.3214})

\bibitem{Greiner:2011mp}
Greiner N, Guffanti A, Reiter T and Reuter J 2011 {\em Phys.Rev.Lett.\/} {\bf
  107} 102002 

\bibitem{Greiner:2012im}
Greiner N, Heinrich G, Mastrolia P, Ossola G, Reiter T {\em et~al.\/} 2012 {\em
  Phys.Lett.\/} {\bf B713} 277--283

\bibitem{vanDeurzen:2013rv}
van Deurzen H, Greiner N, Luisoni G, Mastrolia P, Mirabella E {\em et~al.\/}
  2013 {\em Phys.Lett.\/} {\bf B721} 74--81 
  
\bibitem{Gehrmann:2013aga}
Gehrmann T, Greiner N and Heinrich G 2013 {\em JHEP\/} {\bf 1306} 058

\bibitem{Gehrmann:2013bga}
Gehrmann T, Greiner N and Heinrich G 2013  (\textit{Preprint}
  \eprint{1308.3660})

\bibitem{Cullen:2013saa}
Cullen G, van Deurzen H, Greiner N, Luisoni G, Mastrolia P {\em et~al.\/} 2013
  {\em Phys.Rev.Lett.\/} {\bf 111} 131801

\bibitem{vanDeurzen:2013xla}
van Deurzen H, Luisoni G, Mastrolia P, Mirabella E, Ossola G {\em et~al.\/}
  2013  (\textit{Preprint} \eprint{1307.8437})

\bibitem{Dolan:2013rja}
Dolan M~J, Englert C, Greiner N and Spannowsky M 2013  (\textit{Preprint}
  \eprint{1310.1084})

\bibitem{Cullen:2012eh}
Cullen G, Greiner N and Heinrich G 2013 {\em Eur.Phys.J.\/} {\bf C73} 2388

\bibitem{Greiner:2013gca}
Greiner N, Heinrich G, Reichel J and von Soden-Fraunhofen J~F 2013
  (\textit{Preprint} \eprint{1308.2194})

\bibitem{Chiesa:2013yma}
Chiesa M, Montagna G, Barze` L, Moretti M, Nicrosini O {\em et~al.\/} 2013 {\em
  Phys.Rev.Lett.\/} {\bf 111} 121801;
Mishra K, Becher T, Barze L, Chiesa M, Dittmaier S {\em et~al.\/} 2013
  (\textit{Preprint} \eprint{1308.1430})

\bibitem{Luisoni:2013cuh}
Luisoni G, Nason P, Oleari C and Tramontano F 2013  (\textit{Preprint}
  \eprint{1306.2542})

\bibitem{Hoeche:2013mua}
Hoeche S, Huang J, Luisoni G, Schoenherr M and Winter J 2013 {\em Phys.Rev.\/}
  {\bf D88} 014040

\bibitem{Nogueira:1991ex}
Nogueira P 1993 {\em J.Comput.Phys.\/} {\bf 105} 279--289

\bibitem{Vermaseren:2000nd}
Vermaseren J~A~M 2000  (\textit{Preprint} \eprint{math-ph/0010025})

\bibitem{Reiter:2009ts}
Reiter T 2010 {\em Comput.Phys.Commun.\/} {\bf 181} 1301--1331

\bibitem{Cullen:2010jv}
Cullen G, Koch-Janusz M and Reiter T 2011 {\em Comput.Phys.Commun.\/} {\bf 182}
  2368--2387

\bibitem{Ossola:2006us}
Ossola G, Papadopoulos C~G and Pittau R 2007 {\em Nucl.Phys.\/} {\bf B763}
  147--169;
Ossola G, Papadopoulos C~G and Pittau R 2007 {\em JHEP\/} {\bf 0707} 085;
Ellis R~K, Giele W~T and Kunszt Z 2008 {\em JHEP\/} {\bf 03} 003;
Ossola G, Papadopoulos C~G and Pittau R 2008 {\em JHEP\/} {\bf 0805} 004;
Mastrolia P, Ossola G, Papadopoulos C and Pittau R 2008 {\em JHEP\/} {\bf 0806}
  030 

\bibitem{Mastrolia:2012bu}
Mastrolia P, Mirabella E and Peraro T 2012 {\em JHEP\/} {\bf 1206} 095

\bibitem{Mastrolia:2010nb}
Mastrolia P, Ossola G, Reiter T and Tramontano F 2010 {\em JHEP\/} {\bf 1008}
  080

\bibitem{Binoth:2008uq}
Binoth T, Guillet J~P, Heinrich G, Pilon E and Reiter T 2009 {\em
  Comput.Phys.Commun.\/} {\bf 180} 2317--2330 

\bibitem{Heinrich:2010ax}
Heinrich G, Ossola G, Reiter T and Tramontano F 2010 {\em JHEP\/} {\bf 1010}
  105 

\bibitem{Cullen:2011kv}
Cullen G, Guillet J, Heinrich G, Kleinschmidt T, Pilon E {\em et~al.\/} 2011
  {\em Comput.Phys.Commun.\/} {\bf 182} 2276--2284 

\bibitem{vanOldenborgh:1990yc}
van Oldenborgh G 1991 {\em Comput.Phys.Commun.\/} {\bf 66} 1--15;
Ellis R~K and Zanderighi G 2008 {\em JHEP\/} {\bf 02} 002

\bibitem{vanHameren:2010cp}
van Hameren A 2011 {\em Comput.Phys.Commun.\/} {\bf 182} 2427--2438


\bibitem{TizianoEPS} Peraro T, {\em presentation at EPS-HEP 2013, in these proceedings}

\bibitem{Christensen:2008py}
Christensen N~D and Duhr C 2009 {\em Comput. Phys. Commun.\/} {\bf 180}
  1614--1641;
Degrande C, Duhr C, Fuks B, Grellscheid D, Mattelaer O {\em et~al.\/} 2011
  (\textit{Preprint} \eprint{1108.2040});
Alloul A, Christensen N~D, Degrande C, Duhr C and Fuks B 2013
  (\textit{Preprint} \eprint{1310.1921})

\bibitem{Semenov:2010qt}
Semenov A 2010  (\textit{Preprint} \eprint{1005.1909})

\bibitem{Kuipers:2012rf}
Kuipers J, Ueda T, Vermaseren J and Vollinga J 2013 {\em Comput.Phys.Commun.\/}
  {\bf 184} 1453--1467

\bibitem{Mastrolia:2012du}
Mastrolia P, Mirabella E, Ossola G, Peraro T and van Deurzen H 2012 {\em PoS\/}
  {\bf LL2012} 028 (\textit{Preprint} \eprint{1209.5678})

\bibitem{Cullen:2013cka}
Cullen G, van Deurzen H, Greiner N, Heinrich G, Luisoni G {\em et~al.\/} 2013
  (\textit{Preprint} \eprint{1309.3741})

\bibitem{Binoth:2010xt}
Binoth T, Boudjema F, Dissertori G, Lazopoulos A, Denner A {\em et~al.\/} 2010
  {\em Comput.Phys.Commun.\/} {\bf 181} 1612--1622;
Alioli S, Badger S, Bellm J, Biedermann B, Boudjema F {\em et~al.\/} 2013
  (\textit{Preprint} \eprint{1308.3462})

\bibitem{Dittmaier:2011ti}
Dittmaier S {\em et~al.\/} (LHC Higgs Cross Section Working Group) 2011
  (\textit{Preprint} \eprint{1101.0593});
Dittmaier S, Dittmaier S, Mariotti C, Passarino G, Tanaka R {\em et~al.\/} 2012
   (\textit{Preprint} \eprint{1201.3084});
Heinemeyer S {\em et~al.\/} (The LHC Higgs Cross Section Working Group) 2013
  (\textit{Preprint} \eprint{1307.1347})

\bibitem{Gleisberg:2008ta}
Gleisberg T, Hoeche S, Krauss F, Schonherr M, Schumann S {\em et~al.\/} 2009
  {\em JHEP\/} {\bf 0902} 007

\bibitem{Frederix:2008hu}
Frederix R, Gehrmann T and Greiner N 2008 {\em JHEP\/} {\bf 0809} 122;
Frederix R, Gehrmann T and Greiner N 2010 {\em JHEP\/} {\bf 1006} 086;
Stelzer T and Long W 1994 {\em Comput.Phys.Commun.\/} {\bf 81} 357--371;
Maltoni F and Stelzer T 2003 {\em JHEP\/} {\bf 0302} 027;
Alwall J, Demin P, de~Visscher S, Frederix R, Herquet M {\em et~al.\/} 2007
  {\em JHEP\/} {\bf 0709} 028 

\bibitem{ArkaniHamed:1998rs}
Arkani-Hamed N, Dimopoulos S and Dvali G 1998 {\em Phys.Lett.\/} {\bf B429}
  263--272;
Antoniadis I, Arkani-Hamed N, Dimopoulos S and Dvali G 1998 {\em Phys.Lett.\/}
  {\bf B436} 257--263

\bibitem{Hamilton:2012rf}
Hamilton K, Nason P, Oleari C and Zanderighi G 2013 {\em JHEP\/} {\bf 1305} 082

\bibitem{Mastrolia:2011pr}
Mastrolia P and Ossola G 2011 {\em JHEP\/} {\bf 1111} 014 (\textit{Preprint}
  \eprint{1107.6041});  Badger S, Frellesvig H and Zhang Y 2012
   {\em JHEP\/} {\bf 1204}  055;
  Zhang Y 2012 {\em JHEP\/} {\bf 1209} 042;
Mastrolia P, Mirabella E, Ossola G and Peraro T 2012 {\em Phys.Lett.\/} {\bf
  B718} 173--177;
  Feng B and Huang R 2013
  {\em JHEP\/} {\bf 1302}  117;
Mastrolia P, Mirabella E, Ossola G and Peraro T 2013 {\em Phys.Rev.\/} {\bf
  D87} 085026;
Mastrolia P, Mirabella E, Ossola G and Peraro T 2013  (\textit{Preprint}
  \eprint{1307.5832});
 Badger S, Frellesvig H and Zhang Y 2013
  (\textit{Preprint} 1310.1051)
\end{thebibliography}
\end{document}